\documentclass[10pt, journal, compsocconf]{IEEEtran}
\usepackage{amsmath}
\usepackage{amsfonts}
\usepackage{array}
\usepackage{amssymb}
\usepackage{graphicx}
\usepackage{url}
\usepackage{multicol}
\usepackage{multirow}
\usepackage{stfloats}
\usepackage{subfigure}
\usepackage[usenames]{color}
\usepackage{booktabs}
\usepackage{changes}
\usepackage{eurosym}

\newcommand{\ie}{{\it i.e.}}

\newcommand{\bi}{\begin{itemize}}
\newcommand{\ei}{\end{itemize}}
\newcommand {\beq}{\begin{equation}}
\newcommand {\eeq}{\end{equation}}
\newcommand {\be}{\begin{enumerate}}
\newcommand {\ee}{\end{enumerate}}

\begin{document}

\title{Understanding the evolution of multimedia content in the Internet through BitTorrent glasses}

\author{\IEEEauthorblockN{Reza Farahbakhsh\IEEEauthorrefmark{2}\IEEEauthorrefmark{1},
\'Angel Cuevas\IEEEauthorrefmark{3},
Rub\'en Cuevas\IEEEauthorrefmark{3},
Roberto Gonz\'alez\IEEEauthorrefmark{3},
No\"{e}l Crespi\IEEEauthorrefmark{2} }

\IEEEauthorblockA{\IEEEauthorrefmark{2}Institut Mines-T\'el\'ecom, T\'el\'ecom SudParis\\
\{reza.farahbakhsh, noel.crespi\}@it-sudparis.eu}
\IEEEauthorrefmark{3}Universidad Carlos III de Madrid\\
\{acrumin,rcuevas,rgonza1\}@it.uc3m.es\\
\IEEEauthorrefmark{1}Corresponding author
}

\maketitle

\begin{abstract}

Today's Internet traffic is mostly dominated by multimedia content and the prediction is that this trend will intensify in the future. Therefore, main Internet players, such as ISPs, content delivery platforms (e.g. Youtube, Bitorrent, Netflix, etc) or CDN operators, need to understand the evolution of multimedia content availability and popularity in order to adapt their infrastructures and resources to satisfy clients requirements while they minimize their costs. This paper presents a thorough analysis on the evolution of multimedia content available in BitTorrent.
Specifically, we analyze the evolution of four relevant metrics across different content categories: content availability, content popularity, content size and user's feedback. 
To this end we leverage a large-scale dataset formed by 4 snapshots collected from the most popular BitTorrent portal, namely The Pirate Bay, between Nov. 2009 and Feb. 2012. Overall our dataset is formed by more than 160k content that attracted more than 185M of download sessions. 

\end{abstract}


\section{Introduction}

In the last years Internet traffic has been mostly dominated by multimedia content \cite{Sandivine}. This has led to the development of new technologies to distribute this content: $(i)$ p2p technologies that allow end-users to share content without the necessity of a dedicated infrastructure, $(ii)$ Cyberlockers that are web-based portals that allow users to both upload and download content, $(iii)$ multimedia content distribution platforms such as YouTube (video), Netflix (TV shows and Movies) or Spotify (music). In addition, in order to reduce the cost and improve the efficiency of the content distribution a new network infrastructure namely Content Delivery Network (CDN) was proposed \cite{CDN_book}. Finally, network operators are becoming content providers as well (e.g. most of them offer their own TV and/or Video-on-Demand services) and,  they have to continuously adapt their network infrastructures in order to efficiently serve the large demand of multimedia content. All these players run their own datacenters to store, process and serve the vast amount of content they offer to their clients. A previous study \cite{Netapp} reveals that 27\% of the power going into a datacenter is consumed by storage. Furthermore, a research report from IBM on datacenters operational efficiency \cite{IBMreport} reveals that the most efficient datacenters perform from four to six times more storage optimization (i.e. data compression, data codification, etc) than other datacenters. Some of the employed techniques in those datacenters are MapReduce or data de-duplication that tries to reduce content redundancy. 

The described scenario, along with the expected steady growth of the traffic associated to multimedia content in the near future \cite{cisco:VNI}, makes interesting to study the evolution of the availability, popularity and size of different types of multimedia content distributed through Internet. Understanding this evolution will help the aforementioned players to adapt their algorithms, infrastructures and resources to meet the needs of their clients and, at the same time, increment their revenues. Furthermore, researchers working in different areas such as storage allocation policies, storage optimization algorithms or content compression techniques would benefit from knowing the trends on the evolution of content availability, popularity and size in order to decide the focus of their future research. 

In this paper, we present a first step to study the evolution of the availability, popularity and size of different types of  multimedia content in the Internet. For this purpose we  use BitTorrent as reference system. We believe that BitTorrent is the most appropriate platform to conduct our study due to the following reasons:

\begin{itemize}
\item  BitTorrent is the application that aggregately contributed more Internet traffic in the last decade \cite{ipoque:2007}\cite{ipoque:2009}.  Recent reports reveal that the weight of BitTorrent in the overall Internet traffic has decreased, however it is still responsible for a representative portion of the overall multimedia traffic in the Internet \cite{Sandivine}. Specifically, this report shows that in 2012 BitTorrent is still the application generating the largest fraction of uplink traffic, in the order of 30\%-40\% in Europe and North America, and aggregately (downstream+upstream traffic) is responsible for more than 10 \% of the traffic in North America and 15\% in Europe.

\item The most popular and recent content (e.g. last Hollywood movies) are typically available in BitTorrent.

\item Other successful platforms such as Netflix, YouTube or Spotify are specialized in a single type of content. Instead, BitTorrent offers a broader catalogue of different types of content (e.g, video, audio, games, etc). Therefore, it allows to perform a comparative study of the evolution of availability, popularity and size across the different types of multimedia content.

\end{itemize}

Our study is based on a large scale dataset collected from the most popular BiTorrent portal, namely The Pirate Bay (TPB),  over a period of more than two years between Nov. 2009 and Feb. 2012. Note that TPB indexes millions of content from the most representative multimedia categories including the most recent and popular ones. Furthermore, TPB receives more than twice daily visits compared to the second most popular BitTorrent portal, according to Alexa ranking\footnote{http://www.alexa.com/}. We have collected 4 snapshots over the defined time window that collectively account for more than 160K content that attracted more than 185M download sessions. This dataset constitutes a representative sample including the required information to perform a meaningful analysis of the evolution of content availability, popularity and size in the Internet over the considered period. Furthermore, we quantify the end-users' feedback activity by means of the number of comments that each content receives.

Our main insights are:

\begin{itemize}

\item Video of different types (Movies, TV Shows, Porn) represents 40-50\% of the overall content and attracts 80\% of the download sessions.

\item The median size of the available content has doubled in a two years period. 

\item High-resolution content has increased five-fold in terms of availability to represent 10\% of the multimedia content and downloads in Feb. 2012.

\item Finally, we have observed that end-users' feedback has been always very limited.

\end{itemize}

\section{BitTorrent Overview}
\label{sec:BitTorrent_overview}

This section presents a brief overview of the functionality of BitTorrent as well as some related works.

\subsection{Background}

There are two separated processes in BitTorrent functionality. On the one hand, we find the process in which a user (publisher) makes a content available, or \textit{publishing phase}. On the other hand, once the content is available end-users (consumers) download it in the \textit{downloading phase}.

In the publishing phase, the publisher generates a .torrent file associated to a content and uploads it in a BitTorrent portal such as The Pirate Bay (TPB).
In addition, the publisher registers the content in one (or more) Tracker(s), which is a server that manages and monitors the swarm (the set of peers sharing a content) associated with a given content. As part of its services, the Tracker keeps track of all the peers (\ie~IP addresses) that share the content and  classifies them either as seeders (which have the full content) or leechers (which have only some pieces of the content). The .torrent file includes (among other information): the IP address of the Tracker (and optionally a list of other backup trackers) that manages the swarm associated to the content, the content size and its name. In addition, major torrent portals like TPB provide a web page for every uploaded content that includes information such as size, category, number of leechers and seeders, content description, users' comments, etc.

In the downloading phase, a BitTorrent client gets the .torrent file associated to the desired content from a BitTorrent Portal (e.g. TPB). That client subsequently sends a request to the Tracker included in the .torrent file. The Tracker replies with: $(i)$ the number of seeders and leechers that are currently connected to the swarm, and $(ii)$ N (typically 50 with a limit of 200) random IP addresses of peers participating in the swarm. Next, the BitTorrent client connects to those peers in order to start receiving pieces of the content (and after getting some pieces serves them to other peers). From time to time, during the downloading process, the BitTorrent client may contact the Tracker to obtain more peers.

We must notice that there exists a Tracker-less mechanism where BitTorrent users can download a file without accessing a Tracker by means of a DHT that provides them with other peers within the swarm of the desired content. Most BitTorrent clients allow using Tracker-based, DHT-based or both mechanisms in parallel to download content. Finally, it is worth to mention that a recent study \cite{Matteo_IFIP} reports that 2/3 of BitTorrent users rely (totally or partially) in Trackers to download BitTorrent files.

\subsection{Related Work}

The success of BitTorrent in the last decade has attracted the attention of the research community that have analyzed and modeled BitTorrent behaviour \cite{Guo05:BitTorrentMeasurement}, characterized the BitTorrent ecosystem \cite{keithRoss:ecosystem}, designed new algorithms to improve its performance \cite{Laoutaris2008:bitmax} and addressed security and privacy issues \cite{choffnes:communities}. In addition, in our previous work \cite{CoNEXT_2010} we study BitTorrent content publishers, classify them and evaluate the socio-economic reasons that motivate these users to make content available in BitTorrent (which in most cases is copyrighted material). Therefore, the technical and socio-economical aspects of BitTorrent have been thoroughly studied.

More related to our work, \cite{BT_popularity} models the content popularity of a large set of torrents. We extend this work by analyzing the popularity of different content categories and also characterize the evolution of other parameters such as content availability and size. Finally, \cite{ipoque:2007}\cite{ipoque:2009} analyze the evolution of the Internet traffic in 8 regions of the world covering the most important applications like web browsing, multimedia streaming, p2p file-sharing, one click hosting, etc, including BitTorrent. In addition, the authors also study the representativeness of different content categories. Although their results are interesting, these studies do not consider regions like North-America or Asia, and monitor less than 200K users in Western Europe. In contrast, our datasets include 100x more users without any regional restriction, thus leading to more accurate results in the evolution of multimedia content in BitTorrent.

\begin{table*}[t]
\scriptsize
  \centering
  \caption{Datasets Description}
    \begin{tabular}{ccccc}
    \toprule
         & \textbf{pb09} & \textbf{pb10} & \textbf{pb11} & \textbf{pb12} \\
    \midrule
    \textbf{Crawling Period} &11/28/09 - 12/18/09  & 04/09/10 - 05/05/10 & 10/21/11 - 12/13/11  &    01/28/12 - 02/12/12       \\
    \textbf{Duration (days)} &21  &  27  &                54    &                   16    \\
    \textbf{Torrents} &                15.8K    &     38.2K    &       72.0K  &            21.0K     \\
    \textbf{Downloads} &                   -      &       95.6M     &      79.0M   &    11.1M      \\
    \bottomrule
    \end{tabular}%
  \label{tab:datasets}%
\end{table*}%

\section{Measurement Methodology}
\label{sec:methodology}
The goal of our measurement process is to collect a large number of contents and the following information for each one of them: $(i)$ the content Category/Subcategory as defined by TPB, $(ii)$ the number of associated download sessions, $(iii)$ the content size, $(iv)$ the number of comments provided by end-users.

Towards this end, we leverage the RSS feed of TPB to detect the availability of any new .torrent file. When a new torrent is detected, in addition to gather its size (from the .torrent file) and Category/Subcategory from TPB, our crawler tool periodically queries the tracker in order to obtain the IP addresses of the participants in the content swarm and always solicits the maximum number of IP addresses (\ie~200) from the Tracker\footnote{From Feb. 2012 The Pirate Bay does not store .torrent files anymore, but provide access to them using magnentlinks. Our tool was modified to work with magnetlinks and it is fully functional.}. To avoid being blacklisted by the Tracker, we issue our queries at the maximum rate that is allowed by the tracker (\ie~1 query every 10 to 15 minutes depending on the tracker load).  Given this constraint, we query the tracker from several geographically-distributed machines so that the aggregated information by all these machines provides an adequate high resolution view of the participating peers (\ie~number of download sessions). We continue to monitor a target swarm until we receive 10 consecutive empty replies from the Tracker. This allows us to capture for each new content its size, Category/Subcategory and the number of associated download sessions. 
Finally, in order to gather the number of comments for a given content, we crawled TPB page of all collected content in June 2012. It must be noted that at that time some of the contents collected by our crawling tool had been removed from TPB, and thus we could not gather their number of comments. 

Using the described methodology we have collected four snapshots of TPB content between Nov. 2009 and Feb. 2012. We refer to them as pb09, pb10, pb11 and pb12 based on the year in which each dataset was collected. Table \ref{tab:datasets} summarizes the main characteristics of these datasets (as it is shown in the table we do not have the number of download sessions for pb09). All the snapshots together contribute more than 160K torrents (\ie~contents) and 185M download sessions. These numbers allow us performing a comprehensive analysis on how the content (and its division into different categories) has evolved over the two years period that separates the four datasets.

Note that our tool only collects those peers indexed by trackers but not those using the DHT. This has no impact on our analysis of the evolution of content availability, content size and users' feedback because these metrics are independent of the mechanism used by the peers to download content. Furthermore, since Tracker-based downloads represent 2/3 of the overall download sessions, the sample of peers included in our dataset is representative  enough to derive meaningful conclusions on the evolution of content popularity across  different multimedia categories.

\section{Content Evolution Analysis}
\label{sec:content_type}

In this section we investigate how the relative weight (in \%) of the different content categories evolve in the period under study. For that, we first classify all the collected contents following the Category/Subcategory schema defined by TPB. Following, we analyze each of them from an availability (portion of content available in each category) and a popularity (portion of downloads for each category) perspective.

\subsection{Content Availability Evolution}
\label{subsec:Publishing_Activity_Evolution}

Table \ref{tab:content_per_category} shows the portion of content available in each Category/Subcategory for pb09, pb10, pb11 and pb12 snapshots. 

VIDEO is the dominant category and doubles, in all the snapshots, the number of contents available in any other category. The VIDEO category shows a very slight increment in its presence between pb09 and pb10 from 39\% to 41\%. It keeps a stable growth to reach 52\% (\ie~at this point there was more video content than the sum of all other categories) of the overall content in pb11, and then it surprisingly shows a considerable drop of 6 percentage points (to 46\%) in the two months separating pb11 and pb12.

We now turn our attention to the PORN category that shows an important increment in its representativeness  during the five months between pb09 and pb10. This increase allows PORN scaling from the 5$^{th}$ category in terms of availability in pb09 (8\%) up to the 2$^{nd}$ position in pb10 accounting for 21\% of the total content. From this moment on, it remained in the  2$^{nd}$ position and maintained its weight, 21\% in pb11 and 23\% in pb12. 

\begin{table}[t]
  \centering
  \scriptsize
	\caption{Proportion of  each content type (portion of available content) by categories/subcategories and datasets (pb09, pb10, pb11 and pb12)}
    \begin{tabular}{lcccc}

    \toprule
    \multicolumn{1}{c}{\textbf{Category}} & \textbf{pb09 (\%)} & \textbf{pb10 (\%)} & \textbf{pb11 (\%)} & \textbf{pb12 (\%)} \\
    \midrule
    \textbf{AUDIO} & \textbf{    15.958   } & \textbf{    15.208   } & \textbf{    12.535   } & \textbf{    13.884   } \\
    Music &     10.118    &     10.796    &       7.984    &       8.414    \\
    Audio Books &       0.376    &       0.728    &       0.579    &       0.608    \\
    Sound Clips &       0.162    &       0.076    &       0.095    &       0.120    \\
    FLAC &       1.757    &       1.218    &       1.894    &       1.910    \\
    Other &       3.546    &       2.390    &       1.984    &       2.833    \\\hline
    \textbf{VIDEO} & \textbf{    39.234   } & \textbf{    41.266   } & \textbf{    52.260   } & \textbf{    46.272   } \\
    Movies &     23.004    &     20.084    &     20.623    &     19.924    \\
    Movies DVDR &              -      &       1.625    &       1.448    &       2.029    \\
    Music Videos &       1.646    &       2.340    &       1.151    &       1.608    \\
    Movie Clips &              -      &       0.433    &       0.237    &       0.493    \\
    TV shows &     11.913    &     14.216    &     21.996    &     15.435    \\
    Handhled &       0.207    &       0.258    &       0.353    &       0.110    \\
    Highres - Movies &       1.348    &       0.644    &       1.842    &       1.728    \\
    Highres - TV shows &              -      &       0.603    &       3.690    &       4.039    \\
    3D   &              -      &              -      &       0.072    &       0.014    \\
    Other &       1.115    &       1.062    &       0.849    &       0.890    \\\hline
    \textbf{APPLICATIONS} & \textbf{    16.788   } & \textbf{      9.922   } & \textbf{      3.986   } & \textbf{      5.006   } \\
    Windows &     13.514    &       9.283    &       3.371    &       3.647    \\
    Mac  &       0.726    &       0.258    &       0.238    &       0.345    \\
    UNIX &       0.071    &       0.089    &       0.136    &       0.235    \\
    Handheld &       0.292    &       0.133    &       0.031    &       0.014    \\
    IOS(Ipad/Iphone) &              -      &              -      &       0.051    &       0.302    \\
    Android &              -      &              -      &       0.097    &       0.349    \\
    Other OS &       2.184    &       0.159    &       0.061    &       0.115    \\\hline
    \textbf{GAMES} & \textbf{      4.997   } & \textbf{      3.253   } & \textbf{      3.084   } & \textbf{      4.236   } \\
    PC   &       3.636    &       2.599    &       2.642    &       3.039    \\
    Mac  &       0.039    &       0.037    &       0.043    &       0.072    \\
    PSx  &       0.181    &       0.063    &       0.088    &       0.254    \\
    XBOX360 &       0.201    &       0.099    &       0.070    &       0.148    \\
    Wii  &       0.389    &       0.198    &       0.141    &       0.168    \\
    Handheld &       0.551    &       0.258    &       0.102    &       0.053    \\
    IOS(Ipad/Iphone) &              -      &              -      &       0.026    &       0.211    \\
    Android &              -      &              -      &       0.232    &       0.177    \\
    Other &       0.402    &       0.279    &       0.092    &       0.115    \\\hline
    \textbf{PORN} & \textbf{      8.264   } & \textbf{    21.553   } & \textbf{    21.140   } & \textbf{    23.007   } \\
    Movies &       5.950    &     10.767    &       9.097    &     10.386    \\
    Movies DVDR &              -      &       0.532    &       0.014    &       0.057    \\
    Pictures &       1.232    &       1.688    &       0.971    &       1.206    \\
    Games &       0.091    &       0.026    &       0.015    &       0.077    \\
    Highres - Movies &       0.201    &       0.511    &       1.878    &       2.422    \\
    Movie Clips &              -      &       7.308    &       8.670    &       8.313    \\
    Other &       0.791    &       0.720    &       0.494    &       0.546    \\\hline
    \textbf{OTHER} & \textbf{    14.759   } & \textbf{      8.798   } & \textbf{      6.994   } & \textbf{      7.595   } \\
    E-books &       5.185    &       4.352    &       3.865    &       5.068    \\
    Comics &       0.421    &       1.059    &       1.316    &       1.278    \\
    Pictures &       2.930    &       2.173    &       1.227    &       1.163    \\
    Covers &       0.058    &       0.016    &       0.021    &       0.005    \\
    Physibles &              -      &              -      &              -      &       0.005    \\
    Other &       6.164    &       1.198    &       0.565    &       0.077    \\
    \bottomrule
    \end{tabular}%

    \label{tab:content_per_category}%
\end{table}%

The remaining categories (AUDIO, APPLICATIONS, GAMES and OTHER) follow a common trend. They steadily reduce their weight between pb09 and pb11 and change this slope between pb11 and pb12. Although the trend is similar we can find a much more marked representativeness loss in the APPLICATIONS and OTHER categories. The APPLICATIONS category almost halves its presence between pb09  (16.8\%) and pb10 (10\%), and maintains that decrement to only account for 4\% of the content in pb11, followed by a small increase up to 5\% in pb12. The OTHER category shows a strong decrement of its weight between pb09 (15\%) and pb10 (8.7\%) to later slow down the slope of this loss to end up in 7\% of the total content in pb11 and slightly increases this value (7.5\%) in pb12. Contrary to these cases, GAMES and AUDIO categories present a smoother reduction in their contribution between pb09 and pb11 of 3 percentage points for AUDIO and 2 percentage points for GAMES, to later increase 1 percentage point in pb12.

After analyzing the evolution of each category we can present three interest insights:
\begin{itemize}
\item Movies and TV Shows (in the VIDEO category) are the most available contents. Both subcategories together always sum up more than 34\% of the total content, and they reach a peak of presence in pb11 when both together surpassed 40\%. Furthermore, if we add the PORN-Movies subcategory, we end up with a range between 40\%-50\% for Movies and TV Shows.

\item There is a relevant increment of the High Resolution content. While that type of content only represented about 1.5\% in pb09 and pb10 (summing up Highres-Movies from PORN and VIDEO and Highres-TV Shows from VIDEO), it grew to 7.4\% and 8.2\% in pb11 and pb12, respectively. This increment is due to the irruption of different HD players in the market in the recent years like blue-ray players, Full-HD TVs, Full-HD screens, etc, that have led content providers to make available more High-resolution content to satisfy the demand of end users.

\item The presence of Windows related content has dramati- cally decreased. It represented 13\% of the total available content in pb09, while in the most recent snapshots its presence is reduced to a mere 3\%.

\end{itemize}

\subsection{Content Popularity Evolution}

The previous subsection has analyzed the content availability in TPB. In this subsection, we study the evolution of the popularity of different Categories/Subcategories over time based on the number of download sessions associated with each content in our snapshots.

Table \ref{tab:downloads_per_category} shows the portion of download sessions in each Category/Subcategory for pb10, pb11 and pb12 snapshots. As we mentioned earlier, we did not collect download information for pb09.

VIDEO is the most popular category  attracting more than 3/5 of the downloads in all the snapshots. However, it shows a relevant drop in its popularity over time. VIDEO represented 71\% of the downloads in pb10  and steadily decreased after that, to 64\% and 59\% in pb11 and pb12 respectively. This loss of popularity might be due to the presence of popular Video-on-demand systems offered by content providers and ISPs that have attracted quite a lot of users. It is particularly relevant the case of NetFlix in NorthAmerica that currently is the application generating more downstream traffic \cite{Sandivine}.

PORN appears as the second most popular category among BitTorrent users. Contrary to VIDEO, PORN presents a steady increase in its weight since it accounts for 17\% of the download sessions in pb10, 24\% in pb11 and 31\% in pb12. The growth in the PORN's share (14 percentage points) almost matches the VIDEO category drop (12 percentage points). Finally, it is very important to notice that the sum of these two categories represents about 90\% of the total downloads for the three snapshots. More interestingly, by zooming in our analysis into subcategories, we realize that out of that 90\%, 80\% belongs to the following subcategories: VIDEO/Movies, VIDEO/TV Shows, VIDEO/Highres-Movie, VIDEO/Highres-TV Shows, PORN/Movies, PORN/Highres-Movies.

\begin{table}[t]
  \centering
  \scriptsize
   \caption{Distribution of content popularity (proportion of download sessions) by categories/subcategories and datasets (pb09, pb10, pb11 and pb12)}
    \begin{tabular}{lccc}

      \toprule
    \multicolumn{1}{c}{\textbf{Categories}} & \textbf{pb10 (\%)} & \textbf{pb11 (\%)} & \textbf{pb12 (\%)} \\
    \midrule
    \textbf{AUDIO} & \textbf{      4.671   } & \textbf{      5.574   } & \textbf{      4.972   } \\
    Music &       3.814    &       3.977    &       1.036    \\
    Audio Books &       0.119    &       0.213    &       0.093    \\
    Sound Clips &       0.011    &       0.065    &       0.053    \\
    FLAC &       0.208    &       0.297    &       0.292    \\
    Other &       0.518    &       1.021    &       3.498    \\\hline
    \textbf{VIDEO} & \textbf{    71.299   } & \textbf{    64.080   } & \textbf{    58.925   } \\
    Movies &     41.394    &     29.874    &     22.667    \\
    Movies DVDR &       0.937    &       1.027    &       0.943    \\
    Music Videos &       0.443    &       0.245    &       0.284    \\
    Movie Clips &       0.066    &       0.037    &       0.097    \\
    TV shows &     26.448    &     27.010    &     28.349    \\
    Handhled &       0.127    &       0.040    &       0.014    \\
    Highres - Movies &       0.766    &       3.533    &       3.702    \\
    Highres - TV shows &       0.723    &       2.205    &       2.826    \\
    3D   &              -      &       0.025    &       0.000    \\
    Other &       0.396    &       0.086    &       0.043    \\\hline
    \textbf{APPLICATIONS} & \textbf{      2.117   } & \textbf{      0.996   } & \textbf{      0.810   } \\
    Windows &       2.041    &       0.934    &       0.725    \\
    Mac  &       0.050    &       0.041    &       0.027    \\
    UNIX &       0.002    &       0.002    &       0.000    \\
    Handheld &       0.018    &       0.001    &       0.000    \\
    IOS(Ipad/Iphone) &              -      &       0.003    &       0.002    \\
    Android &              -      &       0.012    &       0.054    \\
    Other OS &       0.006    &       0.001    &       0.001    \\\hline
    \textbf{GAMES} & \textbf{      1.274   } & \textbf{      2.182   } & \textbf{      1.013   } \\
    PC   &       0.790    &       1.747    &       0.756    \\
    Mac  &       0.003    &       0.003    &       0.000    \\
    PSx  &       0.018    &       0.023    &       0.006    \\
    XBOX360 &       0.027    &       0.119    &       0.165    \\
    Wii  &       0.144    &       0.102    &       0.019    \\
    Handheld &       0.216    &       0.022    &       0.001    \\
    IOS(Ipad/Iphone) &              -      &       0.005    &       0.006    \\
    Android &              -      &       0.154    &       0.056    \\
    Other &       0.075    &       0.007    &       0.004    \\\hline
    \textbf{PORN} & \textbf{    17.256   } & \textbf{    24.300   } & \textbf{    31.012   } \\
    Movies &     11.259    &     13.209    &     17.685    \\
    Movies DVDR &       0.034    &       0.014    &       0.025    \\
    Pictures &       0.740    &       0.255    &       0.598    \\
    Games &       0.007    &       0.004    &       0.009    \\
    Highres - Movies &       0.385    &       1.727    &       3.089    \\
    Movie Clips &       4.559    &       8.827    &       8.388    \\
    Other &       0.272    &       0.264    &       1.218    \\\hline
    \textbf{OTHER} & \textbf{      3.383   } & \textbf{      2.868   } & \textbf{      3.268   } \\
    E-books &       1.337    &       2.099    &       2.604    \\
    Comics &       0.326    &       0.225    &       0.115    \\
    Pictures &       1.307    &       0.266    &       0.258    \\
    Covers &       0.003    &       0.000    &       0.000    \\
    Physibles &              -      &              -      &       0.000    \\
    Other &       0.410    &       0.278    &       0.291    \\
    \bottomrule

    \end{tabular}%
   \label{tab:downloads_per_category}%
\end{table}%

The popularity of High-resolution PORN and VIDEO content follows the increasing trend in the availability of this type of content. While High-resolution content only attracted 1.87\% of the downloads in pb10, it has increased its popularity 5 times by receiving 9.62\% of the downloads in pb12.

If we analyze the remaining categories: (i) we find that AUDIO contributes 5\% of the downloads (with variations smaller than 1 percentage point over the three snapshots). (ii) APPLICATIONS goes from 2\% in pb10 to less than 1\% in pb11 and pb12. It is worth noting that APPLICATIONS category contribution is mainly due to Windows applications. (iii) GAMES starts at 1.2\% in pb10, gains 1 percentage point in pb11, and loses it again in pb12. (iv) Finally, the OTHER category remains stable around 3\% with variations smaller than 0.5 percentage points.

In a nutshell, PORN is compensating for loss in VIDEO, which in the worst case attracts 3/5 of the downloads. Both categories together account for 90\% of the downloads. Furthermore, we observe a significant increase in the High-resolution content. Finally,  the rest of the categories remain steady over time with very small variations showing a small but stable interest from BitTorrent consumers in each one of them.

\subsection{Content Availability Vs Content Popularity Discussion}

The most significant content in BitTorrent (according to its major portal, TPB) in terms of availability and popularity are Movies (including porn ones) and TV Shows. Although this type of content represents only 1/2 of the available content, it accounts for 4/5  of the downloads. 

In the case of PORN content we perceive a stable availability (a bit higher than 20\%), but an increment of its popularity, from 17\% to 31\% of download sessions. In particular, PORN is taking up the popularity reduction suffered by the VIDEO category. Similarly to VIDEO, the proportion of available content for PORN is lower than its weight in number of downloads (except in pb10). 

For the rest of the categories the portion of available content exceeds the portion of downloads. The AUDIO category represents between 12\%-15\% of the available content but only attracts 5\% of the downloads. This difference between availability and popularity is due to the presence of several platforms where high quality audio content can be accessed either free (e.g. Spotify) or at low rates (e.g. iTunes). In the case of the GAMES category, it contributes between 3\%-5\% of the content to get 1\%-2\% of the downloads. The OTHER category feeds 7\%-9\% of the content (without considering pb09) and only captures 3\% of the downloads. Finally, the APPLICATIONS category contributes 10\%, 4\% and 5\% of the content in pb10, pb11 and pb12, to attract 2\%, 1\% and 0.8\% of the downloads, respectively.

Therefore, we can conclude that if TPB removes all the categories except VIDEO and PORN, although it would lose half of its available content, it would not suffer a significant reduction in the downloading activity. In addition, the results suggest that High-resolution content is rapidly increasing its availability and popularity.

\section{BitTorrent's Content Size Analysis}
\label{sec:Content_Size_Analysis}

\begin{figure}[t]
\centering
\includegraphics[width=3.8in]{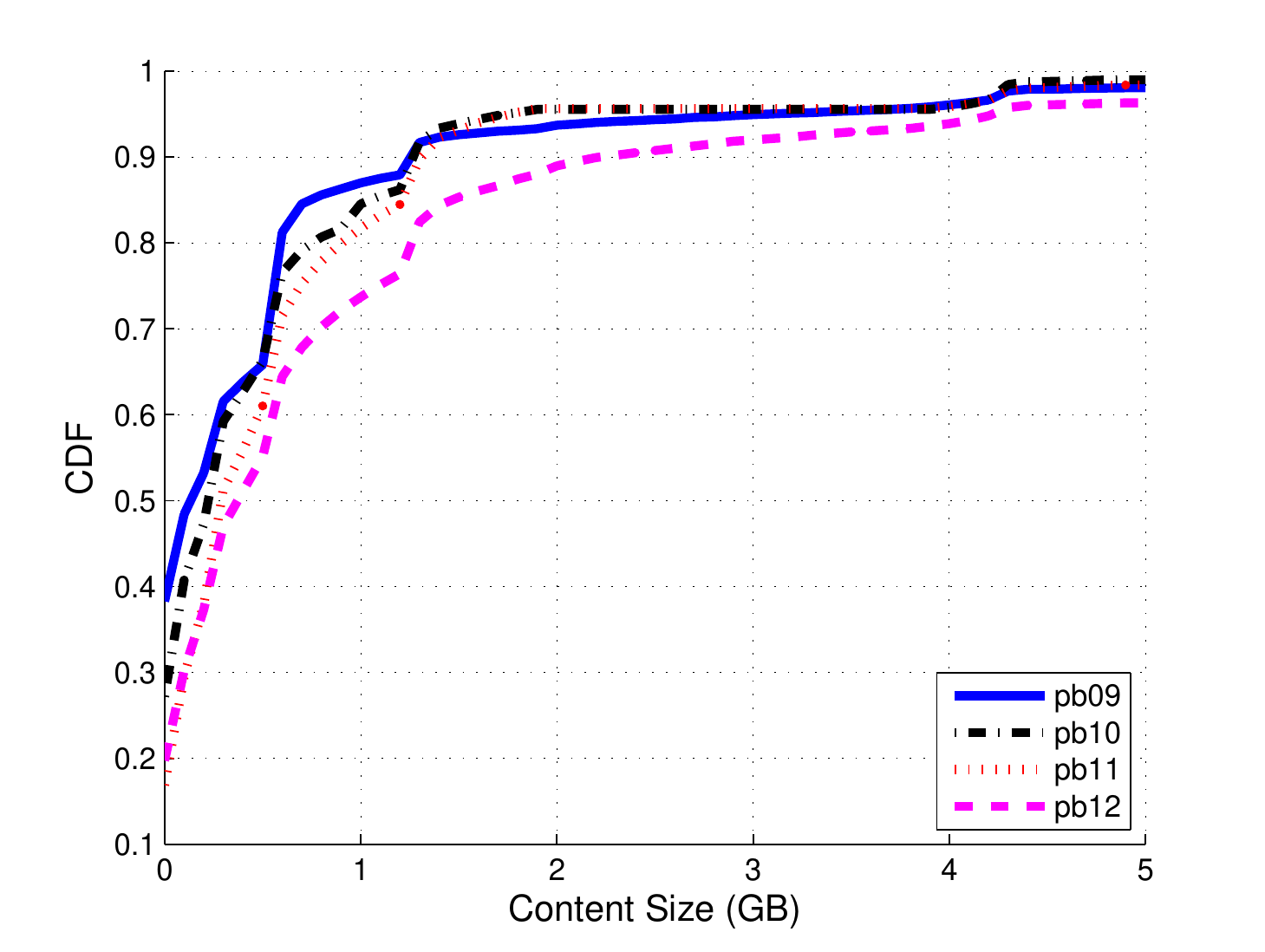}
\caption{Torrents Size CDF}
\label{fig:size_CDF}
\end{figure}

In this section we characterize the evolution of the content size across the four snapshots. This allows us understanding whether the size of BitTorrent content is increasing linked to the presence of everyday larger content in the multimedia arena (e.g increment of High-resolution content presence). To perform this discussion, we will first look at the  aggregate content size distribution across all the snapshots to later narrow down our analysis to individual categories. 

\subsection{Aggregate Content Size Distribution}
Figure \ref{fig:size_CDF} depicts the CDF of content size for our four snapshots. For a better understanding, the graph only shows the CDF for content up to 5 GB (that includes the DVD standard size of 4.7GB), which accounts for more than 96\% of the content within our dataset. The graph shows a steady increase of the content size over the 2-years period under study. The median value of the content size in pb09 was 223MB and increased by 53\%  (to 341MB) in the next five months (pb10). It kept growing up to 370MB  and 458MB in pb11 and pb12 respectively. The conclusion is that BitTorrent content has doubled its size (in median)  in a period of 2 years.

We also want to highlight that the content larger than a standard DVD of 4.7GB (not included in the graph) increases its representativeness by almost 2 percentage points from 2.06\% in pb09 to 3.85\% in pb12. 

\begin{figure*}[t]
\centering
\subfigure[pb09]{
\includegraphics[width=4.2cm,height=4cm]{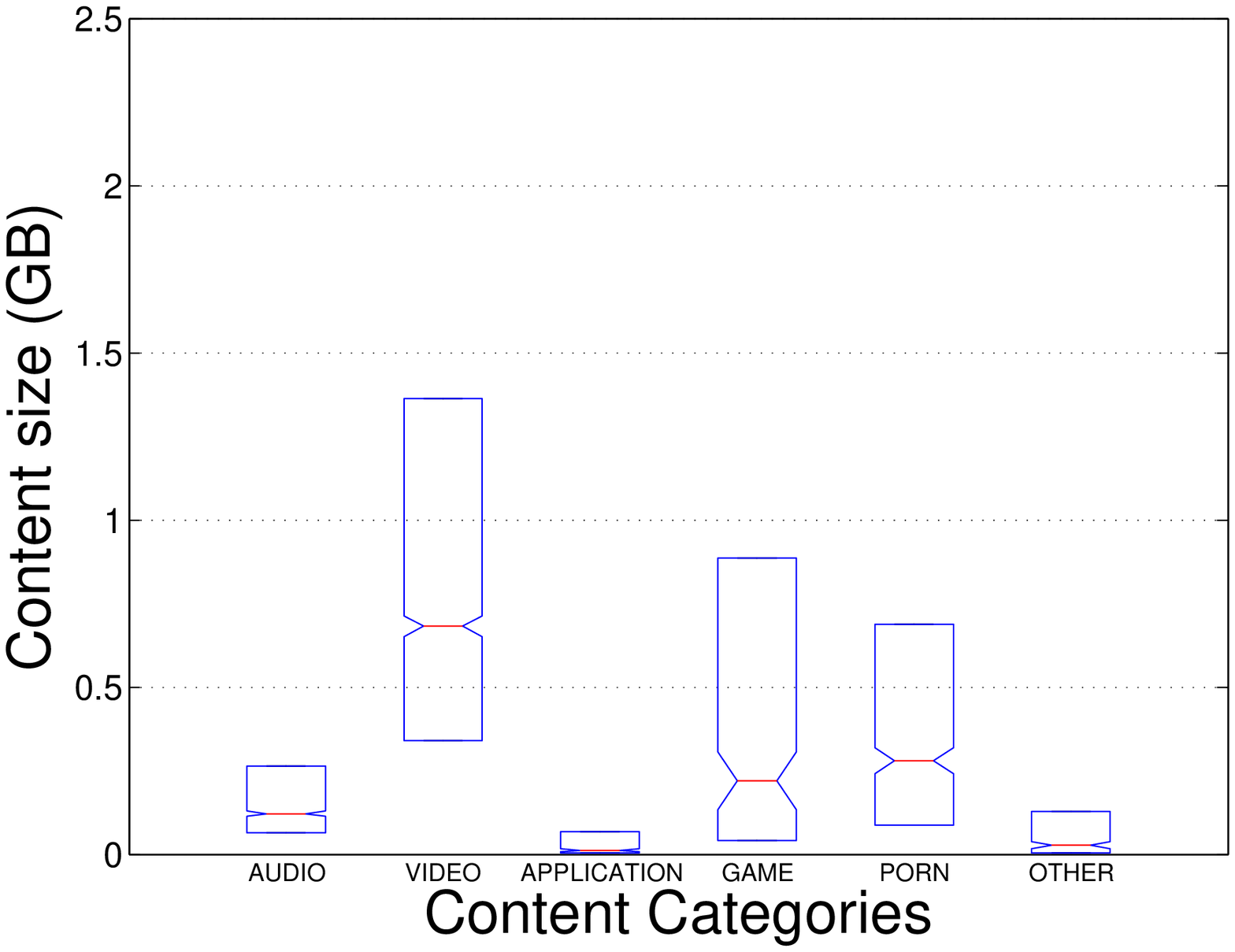}
\label{fig:pb09_size}
}
\subfigure[pb10]{
\includegraphics[width=4.2cm,height=4cm]{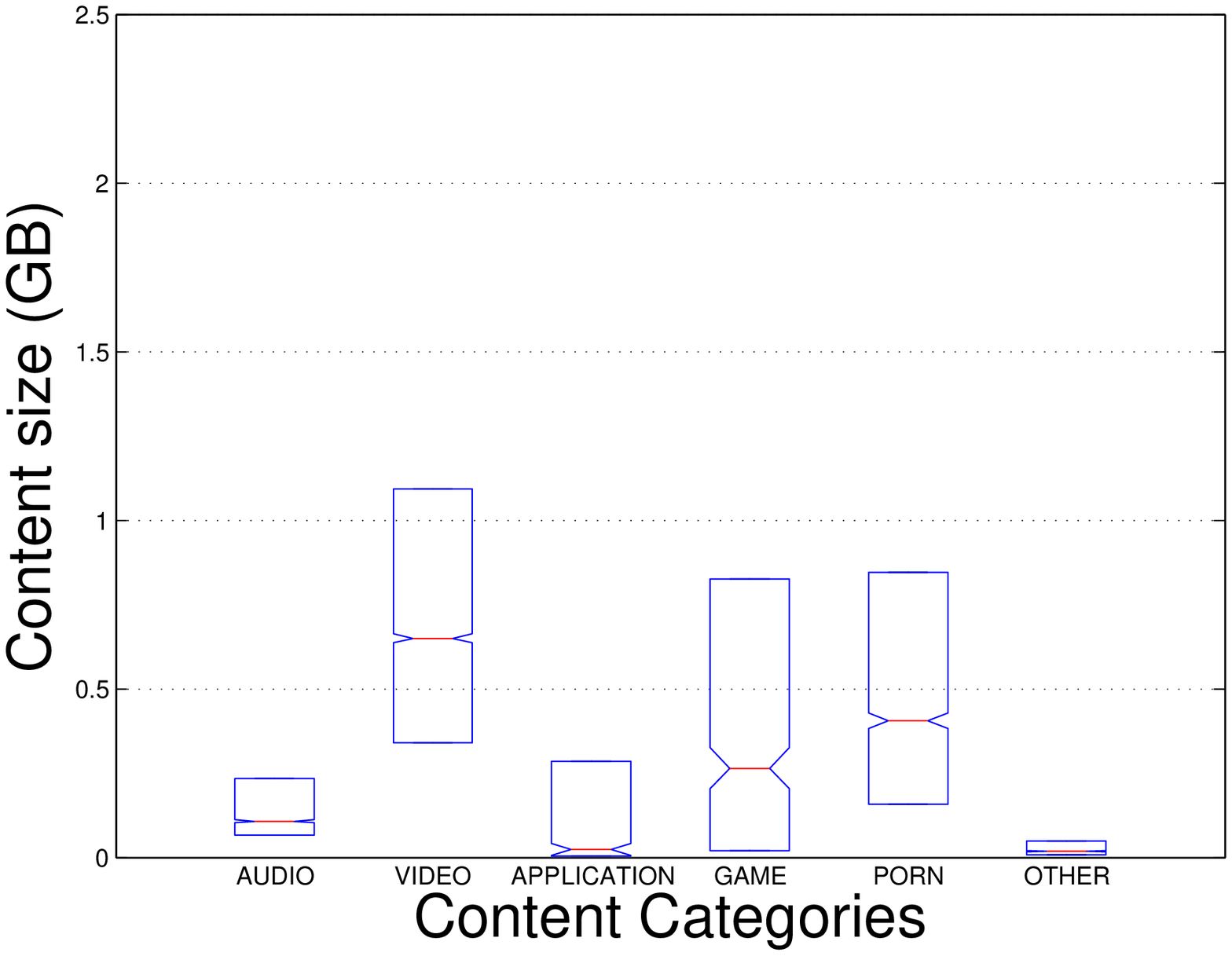}
\label{fig:pb10_size}
}
\subfigure[pb11]{
\includegraphics[width=4.2cm,height=4cm]{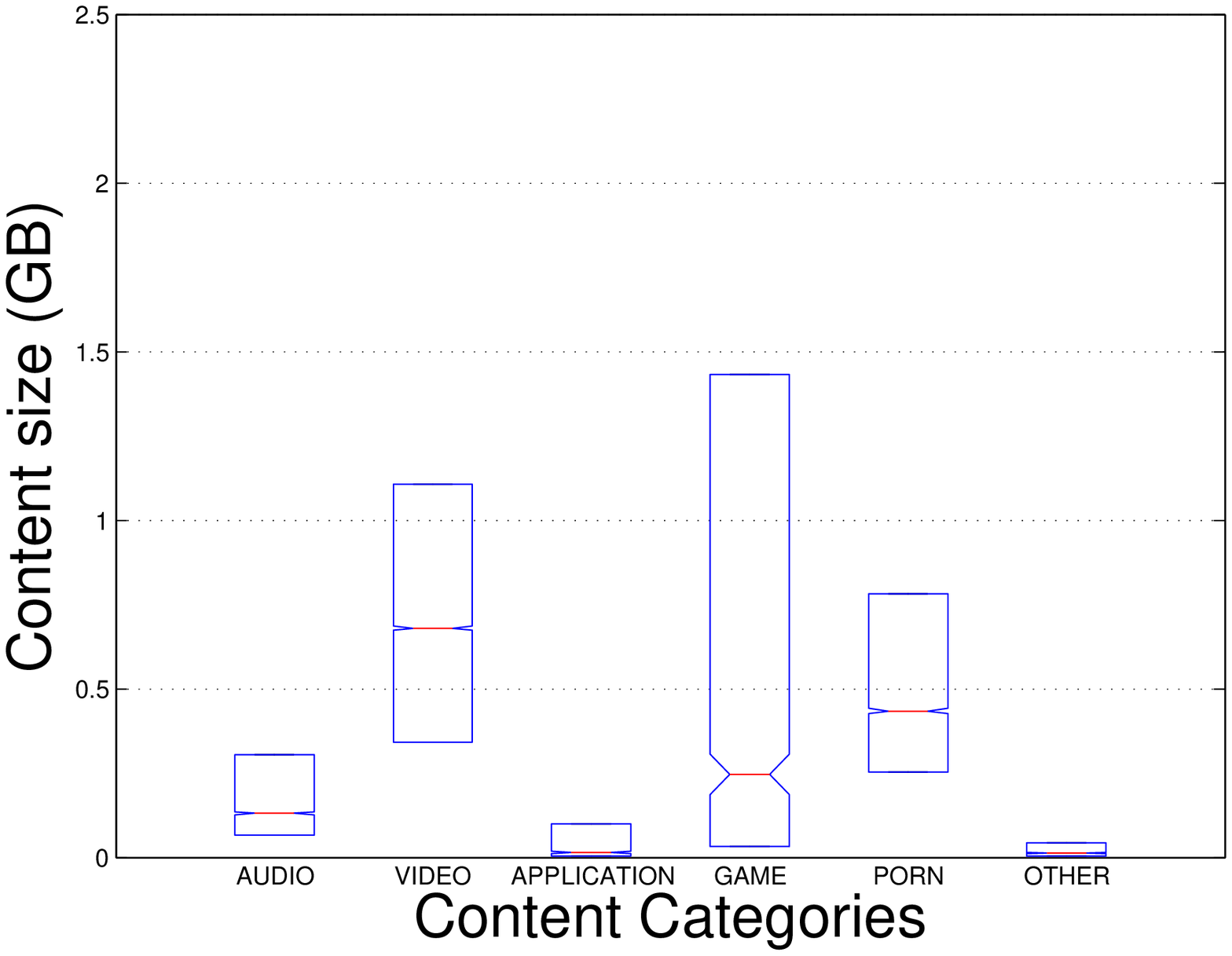}
\label{fig:pb11_size}
}
\subfigure[pb12]{
\includegraphics[width=4.2cm,height=4cm]{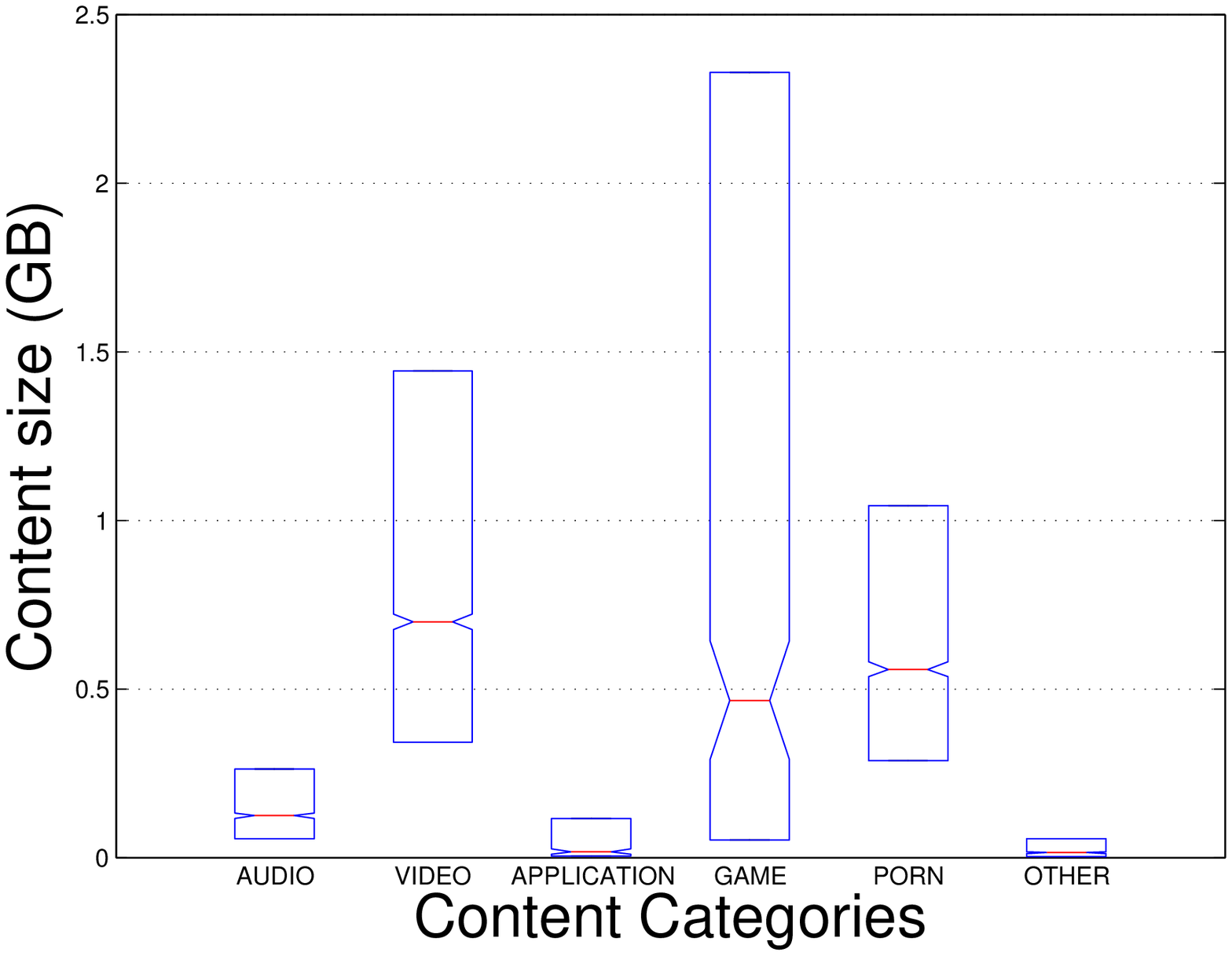}
\label{fig:pb12_size}
}
\caption{Box plot of content size per category for pb09, pb10, pb11 and pb12 datasets. For each category we show the 25th, 50th (median) and 75th percentiles represented by the bottom horizontal blue line, the middle horizontal red line and the top horizontal blue line, respectively.}
\label{fig:size_CDF_per_category}
\end{figure*}

\subsection{Content Size per Category}

Having depicted the overall picture for the content size evolution, we devote our effort to analyze the different categories. Figure \ref{fig:size_CDF_per_category} shows the box plot (which includes 25th-percentile, 50th-percentile or median, and 75th-percentile) of the content size for every category in each one of the four snapshots. The obtained results allow us dividing the categories into two groups: $(i)$ low-size categories composed by AUDIO, APPLICATIONS and OTHER, and $(ii)$ large-size categories formed by VIDEO, GAMES and PORN.

Among the low-size categories, AUDIO is the one presenting a larger size and a very stable distribution over the 2-years period under study. The median size for AUDIO is around 120MB with a small variance. Its 75th-percentile doubles the median and stays close to 250MB, except in pb11 that goes above 300MB. The APPLICATIONS and OTHER categories show very low median values below 25MB in all the datasets. The only remarkable issue for these two categories happens for APPLICATIONS in pb10, which shows a much higher 75th-percentile (285MB)  than in the other cases.

In the large-size categories, VIDEO is the one with the largest median content size over time. It is interesting to observe that it presents a quite stable median value with only a slight variation between 650MB and 700MB. While the 25th-percentile is also stable ($\sim$340MB) in the four snapshots, the 75th-percentile presents a moderate increment of 31\% from its lowest value of 1.1GB in pb10 to the highest one of 1.44GB in pb12. PORN ranks as the category with the second largest median content size. In the case of PORN all percentiles grow over  time. For instance, the median size  evolves as follows: 280MB (pb09), 405MB (pb10), 434MB (pb11) and 558MB (pb12). This demonstrates that PORN content has doubled its median size in a period of only two years. Finally, GAMES is the category that presents a major variance between the different percentiles. The 25th-percentile and the median are always lower than the same parameter in the VIDEO and PORN categories. However, the 75th-percentile becomes the largest one in pb11 and pb12. These large variability between the different percentile thresholds occurs because we can find a large set of games with a very small size (e.g for smartphones, portable videoconsoles, etc), and at the same time a large set of games of very big size (e.g. DVD, Blue-ray, etc). In particular, the extreme variability shown in pb11 and pb12 responds to the recent appearance of video-game consoles and PCs that are Blue-ray capable and the increasing presence of Blue-ray games in the market.

\subsection{Content Size Increment Discussion and Implications}

First of all, it must be noted that those categories that massively contribute to BitTorrent (\ie~VIDEO and PORN) happen to be the ones presenting a larger size, while those categories with a minor presence contain content of small size. The only exception is the GAMES category that shows an extreme variability in the size of its content, especially in pb11 and pb12 snapshots.

The increment of the content size can be explained by 3 main factors: $(i)$ The important evolution in the availability of High-resolution content (large size) which in pb12 already represents 8.2\% of the content. $(ii)$ PORN that represents more than 20\% of the content is doubling its size (in median); and, $(iii)$ the 75th-percentile for VIDEO content size (which represents 40\%-50\% of the available content) has increased a 31\%, probably due to the major presence of High-resolution Movies and TV Shows. 

The fact that BitTorrent content (and by extension mutlimedia content in the Internet) has doubled its size in the last two years is something that major Internet players in the content distribution industry (ISPs, multimedia content providers, Hosting Services, CDN operators, etc) need to take into account in order to update their infrastructures, resources and data processing techniques. For instance, if the content growth speed depicted by our results remains stable over time, these players will need to fairly increase the storage capacity of their datacenters every two years, or perform intensive data processing to avoid such storage capacity increment.

\section{User Comments on BitTorrent Contents}
\label{sec:user_comments}

\begin{table}[htbp]
  \centering
\scriptsize
  \caption{Percentage of contents with comments in different categories (contents with at least 1 comment | contents with three or more comments)}
    \begin{tabular}{|c|c|c|c|c|}
    \hline
       \textbf{ Category} & \textbf{pb09} (\%) & \textbf{pb10} (\%) & \textbf{pb11} (\%) & \textbf{pb12} (\%) \\
        		  & $\geq$1 $|$ $\geq$3 &  $\geq$1 $|$ $\geq$3 & $\geq$1 $|$ $\geq$3 & $\geq$1 $|$ $\geq$3 \\\hline
   \textbf{ AUDIO} & 39.21 $|$ 10.79 & 37.29 $|$ 11.43 & 34.99 $|$ 9.46 & 26.91 $|$ 5.75 \\\hline
    \textbf{VIDEO} & 41.42 $|$ 15.06 & 46.32 $|$ 15.56 & 33.00 $|$ 9.74 & 32.83 $|$ 10.12 \\\hline
    \textbf{APP.} & 47.93 $|$ 19.28 & 72.03 $|$ 31.36 & 61.75 $|$ 25.28 & 53.44 $|$ 20.84 \\\hline
    \textbf{GAMES }& 63.83 $|$ 34.89 & 66.06 $|$ 30.30 & 67.33 $|$ 29.45 & 54.11 $|$ 23.19 \\\hline
    \textbf{PORN} & 33.50 $|$ 6.32 & 31.62 $|$ 6.53 & 15.41 $|$ 2.56 & 12.36 $|$ 2.08 \\\hline
    \textbf{OTHER} & 37.24 $|$ 11.99 & 33.28 $|$ 7.89 & 44.98 $|$ 13.76 & 31.41 $|$ 9.29 \\\hline
   \textbf{Aggregate}& 41.25 $|$ 14.01 & 42.66 $|$ 13.74 & 31.80 $|$ 9.30 & 28.35 $|$ 8.29 \\\hline

    \end{tabular}%
  \label{tab:cat_comments}%
\end{table}%

An interesting aspect related to the BitTorrent content analysis is to study the users' interaction and feedback. In order to measure such activity we have crawled the TPB page of each content in our dataset  (unless they had been removed from TPB) to capture the number of comments that BitTorrent users wrote. With this data we are able to study how the number of comments have evolved over the period under study. 

Table \ref {tab:cat_comments} shows the percentage of aggregate content and per category that received at least one comment on their TPB page as well as the portion of content that collected 3 or more comments. 

First of all, if we look at the aggregate content results we conclude that the social activity around BitTorrent content is quite reduced and the users are just focused on accessing the content without sharing much about its experience. The portion of content receiving three or more contents is in the best case 14\% (pb09) . In addition, we did not find any content with more than nine comments. Furthermore, the number of comments per content decreases over time. For instance the portion of content that presents at least one comment goes down from more than 40\% in pb09 and pb10 to 32\% and 29\% for pb11 and pb12, respectively. This happens because the time is an important variable that increases the likelihood that a content receives one or more comments, the longer the content is exposed the more likely is that a user comments on it.

The GAMES and APPLICATIONS categories are the ones containing a larger portion of content with comments. Although in pb09 GAMES is largely leading this ranking, in the other snapshots both categories present similar results alternating in the first position. We could roughly account between 55\%-65\% of the content with at least one comment and 20\%-30\% with three or more comments. GAMES and APPLICATIONS content usually requires some particular knowledge to manage them (e.g. what movement each button generates in a video game, how to find different options in an application, etc). In addition, in the case of applications the installation process could be challenging for non-skilled users. These two factors increase the need for BitTorrent consumers to interact with the content publisher (or with other consumers of the same content) in order to solve some issue and  manage the downloaded game or application.

After the two leading categories, we can find a second group that includes AUDIO, VIDEO and OTHER categories. We can establish rough intervals of 30\%-40\% and 5\%-15\% for the portion of content in those categories presenting at least one comment and 3 or more comments in their TPB page, respectively. This makes a relevant difference of 15 percentage points from the two previous categories. 

Finally, the PORN category attracts the fewest comments from end-users. Roughly 15\%-30\% of the PORN content shows at least one comment, and only 2\%-6\% has three or more comments. PORN is not only the category with the smallest portion of content attracting comments, but it is also the one experiencing the largest reduction of this parameter over the time. It loses 20 percentage points from 33\% (in pb09) to 12\% (in pb12) during the time period under study. It is obvious that PORN is a very controversial content that is still considered immoral in much of the world, and even forbidden in many countries. Therefore, although it is massively consumed (2nd category in content availability and popularity), consumers prefer not to comment on it.

In a nutshell, this section demonstrates the low interest of BitTorrent users in commenting on their downloaded content. This reveals that users of BitTorrent portals (typically used to distribute copyrighted content) prefer to minimize their visibility, which suggests that they are aware of the illegal nature of their activity.


\section{Conclusion}
\label{sec:conclusion}

This paper has presented a thorough analysis on the evolution of multimedia content available in the most popular BitTorrent portal over a two years period between Nov. 2009 and Feb. 2012. Our results predict a steady and important increment of the multimedia content traffic, which already represents the major part of the Internet traffic, sustained in three main findings: $(i)$ Multimedia content has doubled its size in a period of only 2 years, $(ii)$ the major part (80\%) of the consumed multimedia content corresponds to TV Shows and Movies (including porn) that belong to those categories with the largest size, and $(iii)$ High-resolution content, which has very large size, is increasing its presence and it already represents 8\% of the available content and 10\% of the downloads in our most recent snapshot dated at the beginning of 2012. These findings are useful to those Internet players (\ie~ISPs, content providers, hosting services, CDN operators) involved in the content distribution business in order to update their infrastructures, resources and data processing algorithms to efficiently distribute and serve multimedia content. Furthermore, the significant growth of multimedia content in the last 2 years justifies the research efforts that try to reduce the datacenters storage necessities for the aforementioned players, and thus the results shown in this paper are of high value for those researchers working in that area.


\section*{Acknowledgment}

The research leading to these results was funded by the European Union under the project eCOUSIN (EU-FP7-318398) and the project TWIRL (ITEA2-Call 5-10029), the Spanish MECD under the CRAMNET project (TEC2012-38362-C03-01), the Spanish Ministry of Economy and Competitiveness under the eeCONTENT project (TEC2011-29688-C02-02), and the General Directorate of Universities and Research of the Regional Government of Madrid under the MEDIANET Project (S2009/TIC-1468).





\end{document}